# Community Structure in Interaction Web Service Networks

Chantal Cherifi* and Jean-François Santucci
SPE Laboratory,
Corsica University,
Corte, France
E-mail: chantalbonner@gmail.com
*Corresponding author

**Abstract:** Many real-world complex systems such as social, biological, information as well as technological systems results of a decentralized and unplanned evolution which leads to a common structuration. Irrespective of their origin, these so-called complex networks typically exhibit small-world and scale-free properties. Another common feature is their organisation into communities. In this paper, we introduce models of interaction networks based on the composition process of syntactic and semantic Web services. An extensive experimental study conducted on a benchmark of real Web services shows that these networks possess the typical properties of complex networks (small-world, scale-free). Unlike most social networks, they are not transitive. Using a representative sample of community detection algorithms, a community structuration is revealed. The comparative evaluation of the discovered community structures shows that they are very similar in terms of content. Furthermore, the analysis performed on the community structures and on the communities themselves, leads us to conclude that their topological properties are consistent.

**Biographical notes:** Chantal Cherifi received her Ph.D. degree from Corsica University in Computer Sciences in 2011. She is presently a researcher of SPE laboratory of Corsica University. Her main research interests include complex network and semantic web services.

Jean-François Santucci is Professor in Computer Sciences at the University of Corsica. His main research interest is discrete event modelling and simulation of complex systems. He has been responsible for two European research projects in the area of test and validation of digital systems and in the area of water management processes in Mediterranean islands. He is also working in interdisciplinary research topics associated to archaeology and anthropology.

## 1. Introduction

Many organisations are publishing their core business competencies on the Internet as a collection of Web services. These building blocks for modular applications independent of any software or hardware platform provide a rapid way to share and distribute information between clients, providers and commercial partners. They can be coupled through the Web to create new value-added services during a composition process.

The notion of community is important in the context of Web services. Classically, a community is a set of Web services with similar functionalities grouped together independently of their location, of their maintenance and of their provider. This organisation of the Web services space is primarily intended to increase the efficiency of publication and discovery processes. Communities are also used to sustain high availability of Web services. Rather than aborting a composition because of the eventual unavailability of some Web services at run time, a substitution process enables the use of a different Web service that can perform the same functionality as the failed Web service.

Two approaches are proposed in the literature to group Web services into communities. The first one focuses on organisation model definition. In (Arpinar et al., 2005; Medjahed and Bouguettaya, 2005) communities are "containers" that group a set of Web services that share the same area of interest, they meet the same set of functional requirements. In (Benatallah et al., 2005; Taher et al., 2006) a community is a collection of functionally similar Web services. In (Cherifi, 2011), the author introduce different levels of functional similarities and propose a network representation to group similar Web services. The second one focuses more particularly on automatic classification. In (Bruno et al., 2005; Katakis et al., 2009; Oldham et al., 2004), the proposed solutions rely on supervised learning techniques while in (Azmeh et al., 2008; Konduri and Chan, 2008; Nayak and Lee, 2007), they rely on unsupervised learning techniques.

In this paper, we propose an alternative approach to build Web service communities. Rather than using a similarity relationship, we consider the ability of Web services to be composed to form communities. The notion of community is then different. A community groups Web services that predominantly interact. This is the approach taken in (Dekar and Kheddouci, 2008). The authors propose to gather Web services that frequently interact in the same compositions. Classification takes place on a network of interacting Web services. Network nodes represent the Web services and an interaction link between two Web services is weighted by the number of times the two Web services are composed. The Web services are grouped into clusters using a b-colouring algorithm. To our knowledge, this is the only attempt to structure the Web services space using interaction as an alternative to the notion of the most classical similarity notion. Our work is in this line. We propose to use the complex network framework to investigate the community structure of the Web service composition space.

Indeed, Web services can be seen as an information system that exhibits the two following salient aspects: they are numerous on the Web and they keep complex interaction relationships. Hence, this information system represents a complex system and it can be modelled under the form of networks.

Different network models can be defined according to node and link definitions. Roughly speaking, we can distinguish three types of nodes (Web service, operation or parameter). The links are drawn according to some rules which depend on the Web services description (syntactic or semantic). In this work, we concentrate on syntactic and semantic parameter and operation networks.

Although very promising, few authors have proposed the use of complex networks as a model for the Web services information system. In (Oh, 2006), syntactic network models are presented. Experimental results show that these networks exhibit some of the typical characteristics observed in most real-world networks, such as the small-world and scale-free properties. In this work, we extend the model to semantic Web services. We investigate their community structure. Indeed, we believe that grouping Web services according to interaction relationships is a good alternative to the classical Web service communities based on similarity criteria for the composition process. A comparative analysis on the different partitioning obtained with a set of representative community detection algorithms is performed on a representative benchmark of syntactic and semantic Web services. We also study the topological properties of the communities uncovered by the various community detection algorithms.

The rest of the paper is organized as follows. In section 2, we introduce the models of Web service networks and we describe their design principles. In section 3, we present the selected community detection algorithms as well as the measure for the partitioning comparison and the topological properties of the community structure used in the analysis. Section 4 is devoted to our methodology and experimental results. We present the Web service collection used to build the Web services interaction networks. A comparative analysis of the topological properties of the networks and their community structure is reported. Conclusion and future work are presented in section 5.

## 2. Web service network models

A Web service can be seen from different points of view. It can be considered as a software system that exposes a set of functionalities through its operations. An operation has a set of input and a set of output parameters, i.e. data to be communicated to or from a Web service. Such a view is simply an input/output perspective. Additionally, we can consider preconditions and effects. A precondition defines a set of assertions that must be met before a Web service operation can be invoked. An effect defines the result of invoking an operation. A Web service can also be described by the constraints specification of its operations execution order. In this case, operations which are said to be "equal" when considering the input/output perspective are not if they have different behavioural descriptions. Finally, a set of non functional attributes, like for example the quality of service could be considered. In this paper, we consider a Web service as a distributed application that exports a view of its functionalities in terms of input and output. Hence, a Web service consists of a set of operations with their parameters. Thereafter we use the following notations. A Web service is a set of operations. Its name is represented by a greek letter. Each operation numbered by a digit contains a set of input parameters noted I and a set of output parameters noted O.

Figure 1 represents a Web service α with two operations 1 and 2, input parameters $I_1 = \{a, b\}$, $I_2 = \{c\}$ and output parameters $O_1 = \{d\}$, $O_2 = \{e, f\}$. In a syntactic description, each input or output parameter is described by a string that we designate by *name*. In a semantic description, each parameter is described by an ontological concept that we designate by *concept*.

From this model, we can derive several networks describing the composition process. Indeed, nodes can be either parameters, operations or Web services. When nodes are Web services or operations, links represent an elementary composition between two operations or between two Web services. In the case of parameters, the links represent the dependency relation between input and output parameters. In this paper, we focus on the parameter network and the operation network. In a network whose nodes are parameters, the links represent the operations. In a network whose nodes are operations, the links represent the parameters that allow operations to interact. Although these networks are very different in nature, we call them interaction networks. Indeed, they reflect in different ways the interaction relationships between a set of Web services in a composition process.

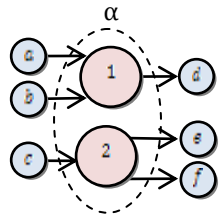

Figure 1. Schematic representation of a Web service α, two operations 1 and 2. $I_1 = \{a, b\}, O_1 = \{d\}, I_2 = \{c\}, O_2 = \{e, f\}$

*2.1 Interaction network of parameters*

An interaction parameter network is defined as a directed graph in which nodes represent the set of parameters and links materialize operations. In other words, a link is created between each of the input parameters of an operation and each of its output parameters. In this context, each operation i can be defined as a triplet *($I_i$, $O_i$, $K_i$)*, where $I_i$ is the set of input parameters, $O_i$ is the set of output parameters and $K_i$ is the set of links dependency. To build the set of interdependencies, we consider that each output parameter of an operation depends on each input parameter of the same operation. The left side of Figure 2 shows three Web services α, β and γ. Their four operations are numbered 1, 2, 3 and 4. The nine input and output parameters are labeled from a to i. As an example of the dependency relationships between the parameters, consider the operation number 2. It is defined by *($I_2$, $O_2$, $K_2$)* where $I_2$ = {c}, $O_2$ = {e, f}, $K_2$ = {(c, e), (c, f)}. Figure 2 (Right side) represents the parameter network corresponding to the three operations on the left. Connectivity within an interaction network of parameters is partly due to the fact that some parameters

can be used by several operations. Moreover, they can be used as input
parameters by some operations and as output parameters by others. For example,
in Figure 2 (Left side), {d, f, g} parameters appear more than once, either as
input or as output of several operations. d is an output of operation 1 and an input
of operation 4, f is an output of 2 and input 3, g is an output of 3 and an input of
4. These parameters are represented by a single node in the network.

*2.2 Interaction network of operations*

An interaction network of operations is defined as a directed graph in which
nodes represent all the Web service operations and relationships materialize an
information flow between operations. Let i be an operation described by its sets
of input and output parameters ($I_i$, $O_i$). To translate an interaction relationship
between this source operation to a target operation j described by the ($I_j$, $O_j$), a
link is created from i to j if and only if for each input parameter of operation j,
there is an similar output parameter of i. In other words, the link exists if
operation i is able to provide all the input data required by operation j. For
example, consider the set of three Web services represented in Figure 2 (Left
side). Web services are named α, β and γ, operations are numbered from 1 to 4,
the input and output parameters are labelled from a to i. The right side
corresponds to the associated interaction network. All entries of operation 3, i.e.,
$I_3 = \{f\}$, are included in the outputs of operation 2, $O_2 = \{e, f\}$, what we translate
by $O_2 \subset I_3$. For this reason, there is a link from operation 2 to operation 3 in the
interaction network (Right side). In this example, no other operation is able to
provide all the input parameters needed by the other operations.

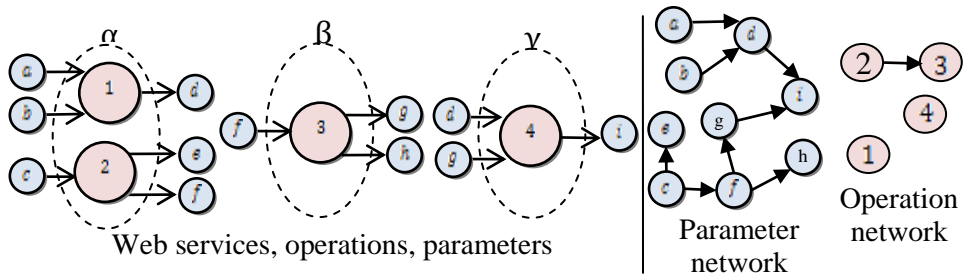

Figure 2. Interaction network of parameters with 9 nodes labelled from a to i and
interaction network of operations with 4 nodes labelled from 1 to 4 (Right side)
obtained from four operations (Left side).

Note that one can use a less restrictive constraint to draw a link between two
operations. We can consider that we can build a composition even if just a subset
of the input parameters needed to invoke an operation is provided.

*2.3 Parameters similarity*

In an interaction network of parameters, we must assess the similarity between two parameters. Similarly, in an interaction network of operations we must decide whether two parameters, one being an output of a source operation and the other being an input of a target operation, are the same. To do so, one must define a matching function. For syntactic described Web services, the matching consists in assessing the similarity between two parameters name. Hence, the syntactic matching consists in measuring the similarity between two strings. Many distance function between strings can be used. In our experiment we consider the strict equality of strings.

In real-world Web services, each provider can use its own naming policies. Therefore, parameters with identical names can convey different information. Similarly, parameters with different names can convey the same information. To tackle this problem, the use of ontological concepts in the semantic descriptions allows more accurate matching. The semantic matching is then based on the notions of similarity between ontological concepts. We compare in this case the concept associated to the parameters. The comparison of two concepts can be achieved by exploring the ontological hierarchy. The matching function can be based on the classical subsumption relationships (Cherifi, 2011). In the following, we only consider networks built with the e*xact* operator. In this case, two parameters are similar if they are described by the same ontological concept.

## 3. Community structure

Many community detection algorithms have been proposed in recent years. Their analysis demonstrates that the objective assessment of the algorithms quality is a complex issue and until now, there is no satisfactory answer to the question of choosing the most appropriate algorithm in the context of a given network and target application. Indeed, depending on the properties of the network, algorithms tend to perform particularly well or particularly poorly. Furthermore, it appears that extrapolating algorithm behaviour on artificial data to real data is not easy. It sometimes leads to contradictory situations, highlighting the structural difference between artificial and real-world networks. Uncovering the community structure of a network therefore cannot rely on a single algorithm. A comparative analysis of outputs of a set of community detection algorithms is more reliable to assess the structural properties of the network. In order to elucidate this issue in our context, we selected a set of community detection methods based on different principles (Fortunato, 2010) among those who received the most attention from the scientific community.

*3.1 Community detection algorithms*

As a representative of the agglomerative hierarchical algorithms, we retain *Louvain* (Blondel et al., 2008) which is more stable than Fast Greedy regarding the nature of the data (Navarro and Cazabet, 2010). It also appears much more efficient than Fast Greedy on artificial data in (Orman, Labatut, and Cherifi, 2011).

For divisive hierarchical algorithms, we retain *EdgeBetweenness* (Girvan and Newman, 2002) that behaves relatively well on artificial data when the community size has the particularity to be heterogeneous (Pons, 2007).

For dynamic random walks based algorithms, we retain *Walktrap* (Pons and Latapy, 2005) and *Infomap* (Rosvall and Bergstrom, 2008) that generally perform well. It is shown that Walktrap behaves relatively well on artificial data in (Pons, 2007) although it performs poorly with small real-world networks (Steinhaeuser and Chawla, 2010). Walktrap and Infomap prove to be the best performing in the comparisons conducted in (Navarro and Cazabet, 2010) and (Orman et al., 2011) on artificial networks.

For the spectral properties based algorithms we use *Eigenvector* (M E J Newman, 2006a). Finally, we retain *Spinglass* (Reichardt and Bornholdt, 2006) as a representative of the simulated annealing optimization approach and *LabelPropagation* (Raghavan, Albert, and Kumara, 2007) which uses the concept of neighbourhood nodes.

*3.2 Comparing two community structures*

Several metrics can be used to measure the similarity between communities delivered by a pair of algorithms. A lot of these measures are strongly correlated (Labatut and Cherifi 2011; Junjie, Hui, and Jian, 2009). We choose to use the normalized mutual information because it is the most commonly used metric in the literature. It has been defined in the context of classical clustering to compare two different partitions of the same data set. Furthermore its interpretation is straightforward. If the community structures are identical, the measure value is 1. If both partitions are independent, the value is 0.

*3.3 Community topological properties*

We can distinguish three types of properties. The global properties are related to the community structure and embody the modularity and the community size distribution. The local properties are related to the communities themselves and encompass the community size, the scaled density, the distance and the hub dominance. Finally one can investigate the "semantic properties", i.e. the operations inside the communities. These properties provide information on how a network is partitioned into communities and on the structural properties of each community.

The *modularity* (Newman and Girvan, 2004) compares the actual proportion of community internal edges to the expected edges proportion if links are randomly distributed. Its value ranges from -1 to 1. For network exhibiting no community structure or when communities are no better than a random partition, the modularity value is negative or equal to 0. In the case of a community structure, the modularity value is between 0 and 1. Practically, a value between 0,3 and 0,7 is considered to be high (M E J Newman 2006b). Modularity's main advantage is that it can be calculated using the network connectivity, in the absence of any node label or other information. Besides, the modularity is very often used as a reference measure in the context of community detection to evaluate the quality of a network partitioning.

The *community size distribution* is an important feature of a community structure. The studies conducted so far on real-world networks tend to show that the community size distribution follows a power law (M E J Newman 2004; Guimerà et al., 2003) with an exponent between 1 and 2. In other words, community size is heterogeneous with the presence of a few large communities and many small ones.

Local properties reveal how well a node is connected to his community and how communities are interconnected. The *density* of a community is defined as the ratio of links it actually contains, to the number of links it could contain if all its nodes were connected. The *scaled density* is a variant obtained by multiplying the density by the community size. When compared to the overall network density, the scaled density allows assessing the cohesion of the community. A community is supposed to be denser than the network it belongs to. If the community is a tree, the scaled density value is 2. If it is a clique, then it is equal to the community size. Some real-world networks such as the Internet or communication networks exhibit tree-like communities. On the contrary, for other classes like social and information networks, the scaled density increases with the community size. Biological networks exhibit hybrid behaviour, their small communities being tree-like, whereas the large ones are denser and close to cliques (Lancichinetti et al., 2010).

The *average distance* of a community can also assess its cohesion. In real-world networks, small communities, smaller than 10, are supposed to have the small world property. So the average distance should increase logarithmically with the size of the community (Lancichinetti et al., 2010). For larger communities, the average distance increases, but more slowly, or stabilizes for certain categories of networks such as communication networks. A small average distance can be explained by a high density in social networks, by the presence of hubs in communication networks and the Internet, or even both in biological networks or information networks.

The *hub dominance* reveals the presence of a central hub in a community. It corresponds to the ratio of the maximal internal degree found in a community, to the maximal degree theoretically possible, given the community size. The hub dominance therefore reaches 1 when at least one node is connected to all others in the community. It can be 0 only if no nodes are connected, which is unlikely for a community. In real-world networks, we observe different behaviours. For communication networks, the value is high in most of communities, independently of their size. This reflects the presence of hubs in all communities. Considering that their structure is sparse and tree form, we can deduce that communities have rather a star structure. This phenomenon is less marked for other types of large real-world networks. One can even notice that the hub dominance decreases while community size increases (Lancichinetti et al., 2010).

# 4. Experimental results

*4.1 Data*

Different benchmarks of publically available Web services description collections are available. They are provided by different entities like the ICEBE organisation (ICEBE'05, 2005), the ASSAM WSDL Annotator project (Hess, Johnston, and Kushmerick, 2004), SemWebCentral (InfoEther and Technologies, 2004), OPOSSum (Küster, König-Ries, and Krug, 2008) or even the authors of (Fan and Kambhampati, 2005). SAWSDL-TC1 (SAWSDL Test Collection 1) is a provided by the SemWebCentral community. We choose to concentrate on this collection for different reasons. First of all, although re-sampled, it is the only one which contains real-world descriptions with both syntactic and semantic information. It allows performing a comparative analysis between the two types of descriptions. Although it is designed to evaluate Web services discovery algorithms, it is a representative sample of Web services that may interact within a composition. Furthermore, in this benchmark, 654 descriptions among the 894 are classified into seven domains. Three of them (economy, education, travel) contain more than 80% of the descriptions. The other four classes (communication, food, medical, weapon) contain less descriptions and are relatively uniform. Moreover, among the highly populated domains, the economy domain seems to be more heterogeneous. It is characterized by the fact that the Web services share the same parameter (or concept) "price". This notion of domain allows us to link the classical notion of communities used in the Web service classification literature, to our alternative approach.

*4.2 Complex Web service networks*

From SAWSDL-TC1 collection, we extracted two parameter networks and two operation networks (one syntactic and one semantic for each node type) using WS-NEXT, a network extractor toolkit specifically designed for this purpose (Cherifi, Rivierre, and Santucci, 2011). The networks are represented on Figure 3, where isolated nodes have been discarded. Globally, all four networks are very similar. A giant component stands next to some small components and isolated nodes. Most real-world networks have such a component structure. The presence of a giant component reflects the fact that the number of possible interactions is high, allowing a large proportion of operations to participate in a composition. As it is commonly done for complex networks exhibiting such a component structure, we focus on the giant component to study the network topological properties (Girvan and Newman, 2002; Oh, 2006).

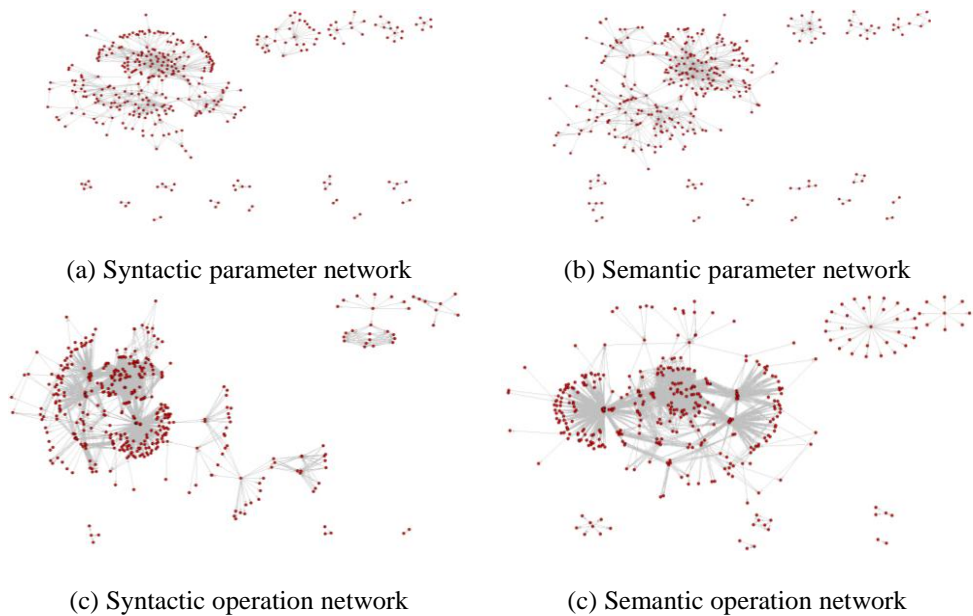

Figure 3. Four Web service networks extracted from SAWSDL-TC1 collection. Syntactic parameter network (a), Semantic parameter network (b), Syntactic operation network (c), Semantic operation network (d).

Networks known as "complex" are generally the result of a decentralized and unplanned evolution. From microscopic mechanisms (i.e. at nodes and links level), such a self-organisation results in the emergence of macroscopic statistical properties. Many complex networks representing very diverse systems share common characteristics. In particular, many of these systems are "small-world". This property popularised by Milgram's experimental study of the structure of social networks, reflects the fact that there is a short path length between any two nodes of the networks. In "small world" networks, the average distance is low and it varies very slowly with the total number of sites (typically as a logarithm). Another important feature is the fact that many networks have few nodes with a very high degree and plenty of nodes with a low degree. This preferential attachment process is commonly illustrated by the expression "the rich get richer". This feature is described by a power law degree distribution and the corresponding networks are said to be "scale-free".

Table 1 shows the measured parameters values allowing to decide if the Web service interaction networks exhibit these two typical properties. All the Web services networks exhibit a smaller average distance as compared to the same size Erdös-Réyni networks and hence have the small-world property. Nevertheless, this property is more pronounced for the semantic networks. This highlights their greater effectiveness for Web service composition. The parameter networks have the scale-free property. To assert this, we conducted Kolmogorov-Smirnov goodness-of-fit tests. Resulting p-values for global degrees are over 0.80, allowing us to conclude that the degree distributions follow a power law distribution. This hypothesis is not confirmed for the operation networks as we observe very low p-values. This may be a consequence of the collection re-sampling. Indeed, we performed the same experiment on the syntactic Public Web Services collection (Fan and Kambhampati, 2005). It appears that for this collection, the power-law degree distribution is a plausible hypothesis with a global degree p-value of 0.79. Though we can consider that interaction Web service networks exhibit the small-world and the scale-free properties.

| Network designation | Network size | Average distance | Average Distance Erdös-Réyni | Power-law exponent | p-value |
|---|---|---|---|---|---|
| Syntactic Parameter | 269 | 2,75 | 6,29 | 3,15 | 0,81 |
| Semantic Parameter | 268 | 1,97 | 6,24 | 3,04 | 0,84 |
| Syntactic Operation | 395 | 2,19 | 3,91 | 2,96 | 0,028 |
| Semantic Operation | 341 | 1.87 | 2,76 | 2, 17 | 0,085 |

Table 1. Topological properties of parameter and operation networks of the SAWSDL-TC1collection: size, average distance, Erdös-Réyni networks average distance, power-law exponent, p-value.

In order to position Web service networks into the complex networks landscape, we recall common topological properties of typical real-world complex networks from various domains, along with values measured on the giant component of the four Web service interaction networks, in Table 2. The considered real-world complex networks are example of information/communication, biological and social networks (Boccaletti et al., 2006). AS2001 and Routers are two information/communication networks. AS2001 stands for the Internet at the autonomous system (AS) level as on April 16th, 2001, while Routers indicate the router level graph representation of the Internet. Gnutella is a peer-to-peer network provided by Clip2 Distributed Search Solutions. Finally, the World Wide Web (WWW) is a directed network formed by the hyperlinks between different Web pages. Each Web page has a number of incoming links and a number of outgoing links pointing to other Web pages. The protein–protein interaction network in the yeast and a network of metabolic reactions are two biological networks. The nodes of the first network are proteins, with two nodes being linked together by an edge if the corresponding proteins physically interact, e.g. if two amino acid chains are binding to each other. Metabolic reaction networks are directed networks whose nodes are chemicals that are connected to one another through the existence of metabolic reactions. Finally, Actors and Math1999 are two examples of social networks. A social network is formally defined as a set of individuals or social entities linked through some kind of interactions among them. Here we consider the collaborations graph of mathematicians defined by paper co-authorships, the movie actor collaboration network based on the Internet Movie Database (a network made up of actors that have casted together in a same movie). Independently from their nature, all the networks exhibit a small average distance and a power-law degree distribution.

High transitivity coefficients are observed in AS2011 and Actors networks. The networks also differ in their degree correlation.

If we look at the small-world property, semantic parameter network and operation networks have an average distance of the magnitude of that of the protein network. The degree distribution of the semantic parameter network is similar to the one of the World Wide Web. We note that the World Wide Web, the metabolic network and the Web services interaction networks are directed networks; the two values of the power-law exponent, respectively represent the in/out-degree exponents of the power-law. The low transitivity of the four networks is similar the one of the router network. Finally, like the network of autonomous systems, Gnutella and biological networks, parameter and operation interaction networks have a disassortative degree correlation.

| Network Designation | Network size | Average distance | Transitivity coefficient | Power-law exponent | Degree correlation |
|---|---|---|---|---|---|
| AS2001 | 11174 | 3.62 | 0.24 | 2.38 | <0 |
| Routers | 228263 | 9.5 | 0.03 | 2.18 | >0 |
| Gnutella | 709 | 4.3 | 0.014 | 2.19 | <0 |
| WWW | $2 \times 10^8$ | 16 | 0.11 | 2.1/2.7 | -- |
| Protein | 2115 | 2.12 | 0.07 | 2.4 | <0 |
| Metabolic | 778 | 7.40 | 0.7 | 2.2/2.1 | <0 |
| Math 1999 | 57516 | 8.46 | 0.15 | 2.47 | >0 |
| Actors | 225226 | 3.65 | 0.79 | 2.3 | >0 |
| Syntactic Parameter | 269 | 2,75 | 0,039 | 3,15/2,01 | <0 |
| Semantic Parameter | 268 | 1,97 | 0.031 | 2.99/3.45 | <0 |
| Syntactic Operation | 395 | 2,19 | 0,032 | -- | <0 |
| Semantic Operation | 341 | 1.87 | 0.022 | -- | <0 |

Table 2. Basic topological properties of Web service networks and of a number of information/communication, biological and social networks from the real-world: vertices number, average distance, transitivity coefficient, exponent of the degree distribution, and type of correlation.

*4.3 Community structure in Web service networks*

Since we do not have any information about the reality of the networks community structure, we proceed by comparison between the partitioning identified by the different algorithms on the different networks. To do this, we have three degrees of freedom: the seven algorithms, the description type (syntactic, semantic) and the network node type (parameter, operation).

We first look at the number of detected communities and then we compare the discovered community content. Table 3 gives the community number for each algorithm and for each network. Statistical trends are described by mean values and standard deviations. Overall, a community structure is revealed in all the cases. In other words, in every situation more than one community is identified.

If we compare the number of communities between a parameter and an operation network using the same description type, it appears that the number of communities is lower in parameter networks than in operation networks. Furthermore, variability of the number of detected communities is much greater in operation networks. We now consider a comparison according to the description type. In other words, we compare parameter networks between them and operation networks between them. In parameter networks, differences are statistically insignificant for the average number of detected communities except for EdgeBetweenness. For all the others, the results are very similar. In operation networks, the average number of detected communities is also very comparable.

If we focus on comparing the different algorithms behaviours, one can notice that in parameter networks, EdgeBetweenness, Louvain, Spinglass and Labelpropagation seem very close in terms of community number. For operation networks, the variability is much greater. One can identify four groups. In an increasing order in terms of community number, we have Eigenvector and Louvain, Spinglass and LabelPropagation, Walktrap and Infomap, and finally Edgebetweness which find eight times more communities than the first group.

|  | PARAMETER NETWORK | | OPERATION NETWORK | |
| --- | --- | --- | --- | --- |
|  | Syntactic | Semantic | Syntactic | Semantic |
| EDGEBETWEENNESS | 9 | 14 | 48 | 43 |
| LOUVAIN | 10 | 10 | 8 | 9 |
| SPINGLASS | 9 | 12 | 12 | 12 |
| EIGENVECTOR | 15 | 12 | 6 | 5 |
| WALKTRAP | 17 | 16 | 23 | 20 |
| INFOMAP | 17 | 18 | 21 | 20 |
| LABELPROPAGATION | 10 | 13 | 9 | 13 |
| Mean | 12,4 | 13,5 | 18,14 | 17,42 |
| Standard deviation | 3,7 | 2,7 | 14,69 | 12,52 |

Table 3. Community number on parameter and operation networks from the SAWSDL-TC1collection, for the community partitioning obtained with seven community detection algorithms.

In order to evaluate the agreement regarding the partitioning content, we compare the partitioning generated by the algorithms in the semantic networks. We use the normalized mutual information measured between partitions identified by the seven algorithms taken two by two. Results are presented in Table 4 under the form of a symmetric matrix which assesses the degree of coherence between the different partitions. Although the number and size of the partitions are generally quite different, it appears that algorithms agree on the contents of these partitions. Eigenvector is the one that stands out most of the others, regardless of network type. LabelPropagation also tends to differ from the others in the operation network. The most consensual algorithms are Spinglass and Walktrap in the parameter network. Indeed, they have the highest (0.85) normalized mutual information average value. In the operation network, Spinglass, Infomap and Louvain are the most in agreement.

|  | Parameter network | | | | | | | Operation network | | | | | | |
|---|---|---|---|---|---|---|---|---|---|---|---|---|---|---|
|  | SPI | WAL | INF | LOU | LAB | EIG | EDG | SPI | WAL | INF | LOU | LAB | EIG | EDG |
| SPI | 1,00 | 0,85 | 0,86 | 0,87 | 0,82 | 0,73 | 0,88 | 1,00 | 0,84 | 0,91 | 0,93 | 0,80 | 0,58 | 0,83 |
| WAL | 0,85 | 1,00 | 0,85 | 0,83 | 0,83 | 0,75 | 0,85 | 0,84 | 1,00 | 0,89 | 0,80 | 0,79 | 0,47 | 0,82 |
| INF | 0,86 | 0,85 | 1,00 | 0,80 | 0,74 | 0,78 | 0,88 | 0,91 | 0,89 | 1,00 | 0,89 | 0,82 | 0,60 | 0,88 |
| LOU | 0,87 | 0,83 | 0,80 | 1,00 | 0,77 | 0,69 | 0,81 | 0,93 | 0,80 | 0,89 | 1,00 | 0,78 | 0,58 | 0,78 |
| LAB | 0,82 | 0,83 | 0,74 | 0,77 | 1,00 | 0,70 | 0,80 | 0,80 | 0,79 | 0,82 | 0,78 | 1,00 | 0,52 | 0,80 |
| EIGE | 0,73 | 0,75 | 0,78 | 0,69 | 0,70 | 1,00 | 0,74 | 0,58 | 0,47 | 0,60 | 0,58 | 0,52 | 1,00 | 0,59 |
| EDG | 0,88 | 0,85 | 0,88 | 0,81 | 0,80 | 0,74 | 1,00 | 0,83 | 0,82 | 0,88 | 0,78 | 0,80 | 0,59 | 1,00 |

Table 4. Normalized mutual information between partitioning in semantic parameter and operation networks. Each box gives the NMI between two partitioning. The name of the algorithms is abbreviated in rows and columns. The i[th] column corresponds to the algorithm presented on the i[th] line.

Figure 4 shows the partitioning obtained by those three algorithms. We effectively can observe the agreement on the community content. In the syntactic networks, we observe similar type of behaviours with some shades. Overall, the consensus is slightly lower (around 5%) in the operation networks than in the parameter networks. In terms of consensus, the ranking of the algorithms remains the same.

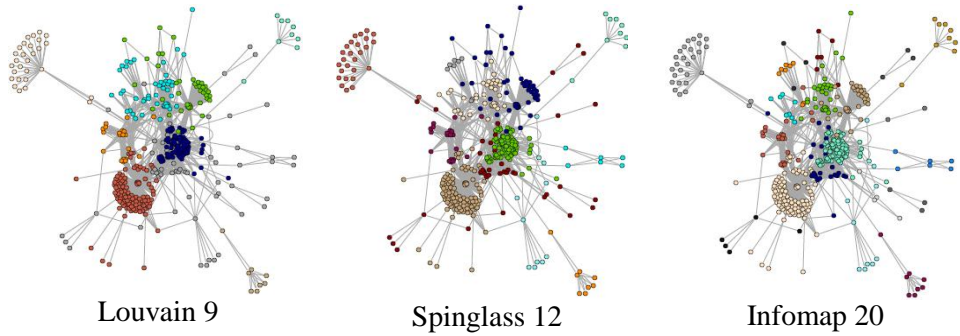

Louvain 9   Spinglass 12   Infomap 20

Figure 4. The three more consensual algorithms in the semantic operation network: Louvain, Spinglass and Infomap with respectively 9, 12 and 20 communities.

*4.4 Communities topological properties*

As we can see in Table 5, the modularity value falls in the range of [0.3, 0.7]. It is higher in parameter networks than in operation networks. In other words, detected communities are more cohesive in parameter networks. If we consider a comparison according to the description type, in parameter networks, differences are statistically insignificant for the average modularity. In operation networks, modularity is systematically higher in the semantic operation network. Comparing the different algorithms behaviours, one can notice that Spinglass is, in all cases, the algorithm that has the highest modularity value. In parameter networks, EdgeBetweenness, Louvain and Spinglass seem very close in terms of modularity. Walktrap, Eigenvector and Infomap are a notch below. For operation networks, Louvain, Spinglass and Infomap have the highest values of modularity.

|  | PARAMETER NETWORK | | OPERATION NETWORK | |
| --- | --- | --- | --- | --- |
|  | Syntactic | Semantic | Syntactic | Semantic |
| EDGEBETWEENNESS | 0,621 | 0,624 | 0,477 | 0,506 |
| LOUVAIN | 0,618 | 0,619 | 0,492 | 0,53 |
| SPINGLASS | 0,637 | 0,63 | 0,508 | 0,53 |
| EIGENVECTOR | 0,6 | 0,596 | 0,434 | 0,479 |
| WALKTRAP | 0,609 | 0,618 | 0,479 | 0,478 |
| INFOMAP | 0,608 | 0,61 | 0,506 | 0,529 |
| LABELPROPAGATION | 0,593 | 0,581 | 0,361 | 0,506 |
| Mean | 0,612 | 0,611 | 0,465 | 0,508 |
| Standard deviation | 0,014 | 0,017 | 0,052 | 0,022 |

Table 5. Community structure modularity on parameter and operation networks from the SAWSDL-TC1collection, for the community partitioning obtained with seven community detection algorithms.

Figure 5 represents the size of the top ten "big" communities discovered by each algorithm in semantic parameter and operation networks. Except for the two biggest communities of the parameter network, the sizes of the communities discovered by the algorithms are quite convergent. Note that Infomap and Eigenvector lead to a more uniform distribution, while LabelPropagation formed the largest community. In the operation network, "opinions" of various algorithms are more convergent with respect to the communities' size, except for Eigenvector which is distinguished by the size of the largest community.

Globally, we observe for all the algorithms quite comparable values for the average distance. Thus, the average distance of the communities in the semantic interaction parameter network ranges from 1 to 2.8. Its mean value is 1.87. It is much smaller than the value observed for the overall network (2.75).

The scaled density values show that small communities tend to be tree like while bigger communities are sparse. Indeed the scaled density ranges from 3 to 5. For this property we observe more variability between the algorithms.

Globally, the hub dominance value is high. It ranges from 0.5 to 0.9. These values suggest the presence of hubs in most communities.

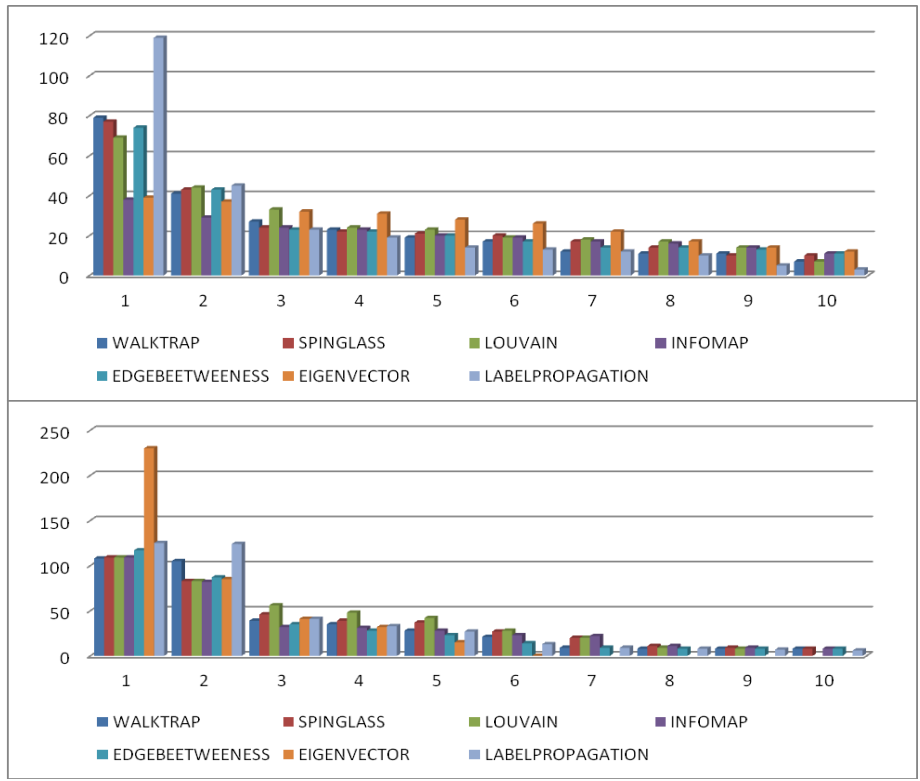

Figure 5. Size of ten largest communities detected by the seven algorithms in semantic parameter (top) and operation (down) networks from SAWSDL-TC1 collection.

We now compare the community's properties distribution between syntactic and semantic networks. For each property we compute the correlation between the data set obtained for the syntactic network and the data set obtained for the semantic network for each algorithm. In all the cases, only the eight larger communities are considered. Indeed, for a given algorithm, the number of detected communities differs between the syntactic case and the semantic case. Furthermore, property values can widely vary when considering small communities. It is preferable to minimize their influence. Table 6 contains the values of the correlation coefficient for each of the algorithms used in parameter networks and operation networks. Overall, we observe large correlation values for all the properties in both types of networks. This suggests that communities found in the syntactic networks are structurally very similar to those found in the semantic networks. Going further into the interpretation of these results is very difficult. Indeed, the differences are difficult to interpret on such a small data set.
We complete this study by conducting a subjective analysis of communities identified in the networks by the different algorithms. For the operation network, we observe that globally, communities and domains do not overlap. Thus, if we focus on the three largest communities, we note that in all the partitioning, they contain a large portion of operations coming from the economy domain with operations from travel and education domains. In medium-sized communities the domain repartition is more homogeneous. In other words, these communities are built with Web services originating from all the domains in comparable proportions. In parameter networks, the community structure is somewhat different. Indeed, in this case, communities are more domain-centred. We explain that by the fact networks are organized around a common vocabulary (the parameters), specific to each domain. From these observations, we can conclude that the notion of community is much richer than the notion of domain. Indeed, a community aggregates services that can potentially enter in a composition, while the notion of domain does not necessarily induce an interaction relationship.

|     | PARAMETER NETWORKS | | | | OPERATION NETWORKS | | | |
| --- | --- | --- | --- | --- | --- | --- | --- | --- |
|     | Size | Distance | Density | Hub Dom. | Size | Distance | Density | Hub Dom. |
| WAL | 0,991 | 0,974 | 0,952 | 0,917 | 0,993 | 0,895 | 0,983 | 0,973 |
| SPI | 0,985 | 0,943 | 0,965 | 0,881 | 0,954 | 0,910 | 0,987 | 0,909 |
| LOU | 0,877 | 0,924 | 0,931 | 0,896 | 0,951 | 0,852 | 0,984 | 0,963 |
| INF | 0,955 | 0,989 | 0,985 | 0,871 | 0,985 | 0,902 | 0,989 | 0,942 |
| EDG | 0,993 | 0,947 | 0,966 | 0,736 | 0,996 | 0,835 | 0,994 | 0,947 |
| EIG | 0,932 | 0,968 | 0,911 | 0,971 | 0,903 | 0,813 | 0,883 | 0,922 |
| LAB | 0,996 | 0,974 | 0,945 | 0,798 | 0,931 | 0,927 | 0,973 | 0,909 |

Table 6. Correlation between the community property' distributions of the syntactic and the semantic networks for each node type. Name of the algorithms is abbreviated.

## 5. Conclusion

In this paper, we explore the topological properties of interaction Web services networks based on syntactic and semantic Web service descriptions. An extensive analysis of four Web service interaction networks extracted from the SAWSDL-TC1 collection is performed. Experimental results show that the networks exhibit the most salient properties characterizing complex networks: the small-world and the scale-free properties. We investigated their community structure, using a set of algorithms based on a broad range of approaches. All the algorithms uncovered a community structure. They mainly agree on the content of the discovered communities. Nevertheless, some differences appear regarding the number of communities. Some algorithms seem to overestimate this number, while others to underestimate it. To better understand the community structure, we analyzed both global and local topological community properties. Measured values of the modularity confirm the community structure of the networks. Indeed, it ranges from 0.36 to 0.62. Those values are currently considered as describing a highly cohesive community structure. Concerning the community size distribution, all the algorithms give similar results except for the two biggest communities where they can behave differently. The average distance value measured in the communities is always lower than the average distance of the overall networks, reflecting the fact that communities are more densely connected. The scaled density value suggests that small communities have a tree structure while bigger communities are sparse. The presence of hubs in the communities is confirmed by the measured hub dominance values which range from 0.5 to 0.9.

The main contributions of this work is to highlight a property of Web service networks which is found in many real-world networks, the community structure, and to validate a new Web service classification approach based on their ability to be composed. Our work is a primer and the approach opens different promising research directions to be investigated. A straight extension is to test and evaluate the proposed classification method for composition synthesis. We conducted preliminary research on a framework based on interaction and similarity networks within which the composition synthesis is guided by the community structure in an interaction network. A future work to be explored is overlapping community detection. So far, the focus in community detection has been on identifying disjoint communities. It is well known that in networks, there could be nodes that belong to more than one community. The overlapping community structure hypothesis can lead to more opportunities in the Web service composition process.

## References

Arpinar, I., Aleman-Meza, B. & Zhang, R., 2005. Ontology-driven web services composition platform. *Inf. Syst. E-Business Management*, vol. 3.

Azmeh, Z. et al., 2008. WSPAB: A Tool for Automatic Classification & Selection of Web Services Using Formal Concept Analysis. In *2008 Sixth European Conference on Web Services*. IEEE, pp. 31-40.

Benatallah, B., Dumas, M. & Sheng, Q.Z., 2005. Facilitating the Rapid Development and Scalable Orchestration of Composite Web Services. *Distributed and Parallel Databases*, 17(1), pp.5-37.

Blondel, V.D. et al., 2008. Fast unfolding of communities in large networks. *Journal of Statistical Mechanics: Theory and Experiment*, 2008(10), p.P10008.

Boccaletti, S. et al., 2006. Complex networks: Structure and dynamics. *Physics Reports*, 424(4-5), pp.175-308.

Bruno, M. et al., 2005. An Approach to support Web Service Classification and Annotation. In *2005 IEEE International Conference on e-Technology, e-Commerce and e-Service*. IEEE, pp. 138-143.

Cherifi, C., 2011. *Classification et Composition de Services Web : Une Perspective Réseaux Complexes*. Corsica University.

Cherifi, C., Rivierre, Y. & Santucci, J.-F., 2011. WS-NEXT, a Web Services Network Extractor Toolkit. In *International Conference on Information Technology (ICIT'11)*. Amman.

Dekar, L. & Kheddouci, H., 2008. A Graph b-Coloring Based Method for Composition-Oriented Web Services Classification. In A. An et al., eds. Berlin, Heidelberg: Springer Berlin Heidelberg, pp. 599-604.

Fan, J. & Kambhampati, S., 2005. A snapshot of public web services. *ACM SIGMOD Record*, 34(1), p.24.

Fortunato, S., 2010. Community detection in graphs. *Physics Reports*, 486(3-5), pp.75-174.

Girvan, M. & Newman, M. E. J., 2002a. Community structure in social and biological networks. *Proceedings of the National Academy of Sciences*, 99(12), p.7821.

Girvan, M. & Newman, M. E. J., 2002b. Community structure in social and biological networks. *Proceedings of the National Academy of Sciences of the United States of America*, 99(12), pp.7821-6.

Guimerà, R. et al., 2003. Self-similar community structure in a network of human interactions. *Physical Review E*, 68(6).

Hess, A., Johnston, E. & Kushmerick, N., 2004. ASSAM: A tool for semi-automatically annotating semantic web services. In *Proc of the 3rd International Semantic Web Conference*.

ICEBE'05, 2005. IEEE International Conference on e-Business Engineering (ICEBE). Available at: http://ieeexplore.ieee.org/xpl/mostRecentIssue.jsp?punumber=10403.

InfoEther & Technologies, B.B.N., 2004. SemWebCentral. , 2011(January). Available at: http://wwwprojects.semwebcentral.org/.

Junjie Wu, Hui Xiong & Jian Chen, 2009. Adapting the right measures for K-means clustering. In *Proceedings of the 15th ACM SIGKDD international conference on Knowledge discovery and data mining*. New York: ACM, pp. 877-886.

Katakis, I. et al., 2009. On the Combination of Textual and Semantic Descriptions for Automated Semantic Web Service Classification. In Boston, MA: Springer US, pp. 95-104.

Konduri, A. & Chan, C.C., 2008. Clustering of Web Services Based on WordNet Semantic Similarity.

Küster, U., König-Ries, B. & Krug, A., 2008. OPOSSum - An Online Portal to Collect and Share SWS Descriptions. In *International Conference on Semantic Computing*. pp. 480-481.

Labatut, V. & Cherifi, H., 2011. Evaluation of Performance Measures for Classifiers Comparison. *Ubiquitous Computing and Communication Journal*, 6, pp.21-34.

Lancichinetti, A. et al., 2010. Characterizing the Community Structure of Complex Networks O. Sporns, ed. *PLoS ONE*, 5(8), p.8.

Medjahed, B. & Bouguettaya, A., 2005. A Dynamic Foundational Architecture for Semantic Web Services. *Distributed and Parallel Databases*, 17(2), pp.179-206.

Navarro, E. & Cazabet, R., 2010. Détection de communautés, étude comparative sur graphes réels. In *MARAMI*.

Nayak, R. & Lee, B., 2007. Web Service Discovery with additional Semantics and Clustering. In *IEEE/WIC/ACM International Conference on Web Intelligence (WI'07)*. IEEE, pp. 555-558.

Newman, M E J, 2004. Detecting community structure in networks. *The European Physical Journal B Condensed Matter*, 38(2), pp.321-330.

Newman, M E J, 2006a. Finding community structure in networks using the eigenvectors of matrices. *Physical Review E - Statistical, Nonlinear and Soft Matter Physics*, 74(3 Pt 2), p.036104.

Newman, M E J, 2006b. Modularity and community structure in networks. *Proceedings of the National Academy of Sciences of the United States of America*, 103(23), pp.8577-8582.

Newman, M. E. J. & Girvan, M., 2004. Finding and evaluating community structure in networks. *Physical Review E*, 69(2), p.26113.

Oh S.-C., 2006. *Effective Web Services Composition in diverse and large-scale services networks*. Pennsylvania State University.

Oldham, N. et al., 2004. METEOR-S Web Service Annotation Framework with Machine Learning Classification. In In Proc. of the 1 st Int. Workshop on Semantic Web Services and Web Process Composition (SWSWPC'04.

Orman, G., Labatut, V. & Cherifi, H., 2011. Qualitative Comparison of Community Detection Algorithms. In H. Cherifi, J. M. Zain, & E. El-Qawasmeh, eds. *Digital Information and Communication Technology and Its Applications*. Springer, pp. 265-279.

Pons, P., 2007. *Détection de communautés dans les grands graphes de terrain*. Paris 7.

Pons, P. & Latapy, M., 2005. Computing communities in large networks using random walks. *Journal of Graph Algorithms and Applications*, 10(2), pp.191-218.

Raghavan, U.N., Albert, R. & Kumara, S., 2007. Near linear time algorithm to detect community structures in large-scale networks. *Physical Review E - Statistical, Nonlinear and Soft Matter Physics*, 76(3 Pt 2), p.036106.

Reichardt, J. & Bornholdt, S., 2006. Statistical mechanics of community detection. *Physical Review E*, 74(1), p.16110.

Rosvall, M. & Bergstrom, C.T., 2008. Maps of random walks on complex networks reveal community structure. *Proceedings of the National Academy of Sciences of the United States of America*, 105(4), pp.1118-1123.

Steinhaeuser, K. & Chawla, N.V., 2010. Identifying and evaluating community structure in complex networks. *Pattern Recognition Letters*, 31(5), pp.413-421.

Taher, Y. et al., 2006. Towards an Approach for Web Services Substitution. In *10th International Database Engineering and Applications Symposium (IDEAS'06)*. IEEE, pp. 166-173.